\documentclass[a4paper,fleqn]{cas-dc}

\usepackage[numbers,sort&compress]{natbib}

\usepackage{graphicx}
\usepackage{dsfont}
\usepackage{subfig}
\usepackage{booktabs}
\usepackage{tabularx}
\usepackage{multirow}
\usepackage{makecell}
\usepackage{array}
\usepackage{threeparttable}
\usepackage{stfloats}
\usepackage{amssymb}
\usepackage{amsmath}
\usepackage{bm} 
\usepackage{siunitx}
\usepackage{lineno}

\usepackage{xcolor}
\definecolor{ElsevierBlue}{HTML}{00AEEF}
\usepackage{hyperref}
\hypersetup{
  colorlinks=true,
  linkcolor=ElsevierBlue,
  citecolor=ElsevierBlue,
  urlcolor=ElsevierBlue,
}

\usepackage{algorithm}
\usepackage{algpseudocode}

\AtBeginDocument{%
}

\begin{document}

\shorttitle{Edge-ANN for Edge Remote Sensing Retrieval} 
\shortauthors{X. Lv et al.}

\title [mode = title]{Edge-ANN: Storage-Efficient Edge-Based Remote Sensing Feature Retrieval}                      

\author[1,2]{Xianwei Lv}
\ead{lvxianwei@neuq.edu.cn}
\author[1]{Debin Tang}
\author[1]{Zhecheng Shi}
\author[3]{Wang Wang}
\cormark[1]
\ead{wangw00821@gmail.com}
\author[1]{Yujiao Zheng}
\author[4]{Xiatian Zhu}

\affiliation[1]{organization={School of Computer and Communication Engineering, Northeastern University at Qinhuangdao},
                addressline={Qinhuangdao},
                city={HeBei},
                postcode={066004}, 
               country={China}}
\affiliation[2]{organization={Hebei Key Laboratory of Marine Perception Network and Data Processing, Northeastern University at Qinhuangdao},
                addressline={Qinhuangdao},
                city={HeBei},
               postcode={066004}, 
                country={China}}
\affiliation[3]{organization={Sustainability X-Lab, The University of Hong Kong},
                addressline={Pokfulam Road}, 
                city={Hong Kong},
                country={China}}
\affiliation[4]{organization={Surrey Institute for People-Centred Artificial Intelligence, and Centre for Vision, Speech and Signal Processing, University of Surrey},
                city={Guildford},
                country={UK}}

\begin{abstract}
Meeting real-time constraints for high-performance Approximate Nearest Neighbor (ANN) search remains a critical challenge in remote sensing edge devices, which are essentially fusion systems like micro-satellites and UAVs, largely due to stringent limitations in primary (RAM) and secondary (disk) storage. To address this challenge, we propose \textbf{Edge-ANN}, an innovative ANN framework specifically engineered for storage efficiency. The core innovation of Edge-ANN lies in its departure from traditional tree-based methods that store high-dimensional hyperplanes. Instead, it leverages pairs of existing data items, termed "anchors," to implicitly define spatial partitions. To ensure these partitions are both balanced and effective, we have developed a novel Binary Anchor Optimization algorithm.This architectural shift eliminates the dimension-dependence of the space complexity. Rigorous experiments on three multi-source datasets, MillionAID, High-resolution Urban Complex Dataset, and GlobalUrbanNet Dataset, demonstrate that under simulated edge environments with dual storage constraints, Edge-ANN achieves a 30-40\% reduction in secondary storage compared to the baseline, at the cost of a minor 3-5\% drop in retrieval accuracy. Furthermore, its overall retrieval performance surpasses that of other mainstream methods in these constrained scenarios. Collectively, these results establish Edge-ANN as a state-of-the-art solution for enabling large-scale, high-performance, real-time remote sensing feature retrieval on edge devices with exceptionally constrained storage. The codes of Edge-ANN are available at \url{https://github.com/huaijiao666/Edge-ANN}.

\end{abstract}


\maketitle

\section{Introduction}
\label{sec:introduction}

Edge-based remote sensing feature retrieval \cite{janakiraman2021study} acts as a pivotal intelligent bridge, connecting vast archives of remote sensing data with real-time emergency response applications. This technology is catalyzing a profound paradigm shift within the remote sensing domain, evolving the conventional "collect-transmit-process" pipeline into a "sense-decide-act" closed-loop intelligence paradigm, executed directly at the sensor front-end \cite{deren2014automatic}. It can equip high-resolution satellites with on-orbit autonomous decision-making, allowing them to selectively transmit only critical imagery of disaster areas from terabytes of raw data to optimize constrained bandwidth \cite{shi2025satellite}. It also permits Unmanned Aerial Vehicles (UAVs) to locate key search-and-rescue targets within minutes during the "golden hour" of post-disaster relief \cite{khankeshizadeh2024novel}. Furthermore, in challenging GPS- or communication-denied environments, this technology bolsters the reliability of autonomous navigation for airborne platforms through visual matching \cite{zhao2019review}. By pre-positioning intelligent decision-making at the sensor level, this technology can deliver unprecedented real-time response capabilities for critical applications like disaster monitoring, predicting grain yield \cite{li2022development,zhou2017predicting,zhang2022expandable,garcia2024advancements}. However, research in remote sensing feature retrieval has predominantly focused on server-side feature extraction and retrieval accuracy, often overlooking the application requirements of storage-constrained edge devices (e.g., the DJI Mini 3 Pro, with only 1GB RAM of main CPU and 512MB RAM of P1-CPU; see Material 1).\par

The pipeline for remote sensing feature retrieval generally consists of two key steps: (1) feature extraction from images and (2) similarity search in the feature space. In the early stages, the feature retrieval was limited to representations based on hand-crafted features, such as SIFT and color histograms \cite{lowe2004distinctive}. The advent of deep learning has shifted mainstream approaches towards employing neural networks for feature extraction. These networks yield highly discriminative deep features, and image similarity is subsequently quantified by the distance between these features \cite{santini2002similarity}. A key challenge lies in the direct adoption of general-purpose similarity search algorithms, which are fundamentally at odds with the strict storage constraints of remote sensing edge devices. Their presumption of abundant resources makes direct deployment problematic.

Fundamentally, remote sensing feature retrieval can be conceptualized as a \textit{k}-nearest neighbor (\textit{k}-NN) search in a high-dimensional feature space \cite{zhu2024cross}. Although brute-force search guarantees globally optimal results, its inefficiency makes it impractical for large-scale, high-dimensional datasets. As a result, approximate \textit{k}-nearest neighbor (ANN) search has emerged as a practical necessity, trading a negligible margin of error for orders-of-magnitude improvement in efficiency, thus becoming the favored solution \cite{zhang2024efficient,heo2018distance,esmaeili2012fast}. Mainstream ANN algorithms can be broadly categorized into three paradigms:1) \textbf{Graph-based methods} like HNSW \cite{malkov2018efficient}, which enable high-precision fast search via hierarchical graphs at the cost of substantial storage for adjacency information;  and 2) \textbf{Tree-based methods}, including KD-Tree \cite{zhou2008real} and ANNoy \cite{annoy2015}, which are simple to maintain but require storing \textit{d}-dimensional hyperplanes at each node, incurring a storage burden proportional to the feature dimensionality; 3) \textbf{Hybrid Graph-and-Tree-Based methods} including DiskANN \cite{jayaram2019diskann}, which leverages a graph to store high-precision neighbor relationships within a tree-like structure, but remaining storage burdens to store hyperplanes.\par

Consequently, deploying ANN on remote sensing edge devices with constrained primary and secondary storage faces two fundamental challenges:
\begin{itemize}
    \item[\textbullet] \textbf{Challenge 1 (Model Lightweighting):} This challenge aims to drastically reduce the model’s storage demands to meet the strict limitations of edge devices, while preserving its core retrieval capabilities.
    \item[\textbullet] \textbf{Challenge 2 (Efficiency and Accuracy Assurance):} This challenge seeks to maintain high retrieval performance (efficiency and accuracy) post-lightweighting, ensuring a viable balance between performance and resource consumption.
\end{itemize}

To address these challenges, we propose \textbf{Edge-ANN}, a storage-optimised ANN framework tailored for edge computing. The key tenet of our approach is to avoid explicitly storing the high-dimensional normal vectors of partitioning hyperplanes in a tree structure. Instead, for each node partition, we repurpose two data points from the dataset as “anchors.” The partitioning hyperplane is defined as the perpendicular bisector of the segment connecting these anchors, which is then adjusted by a learned scalar offset. Consequently, our method only requires storing the two anchors’ fixed-length indices and the offset scalar. This mechanism fundamentally decouples the model size from the feature dimension.

\begin{figure*}[t!]
    \centering
    \includegraphics[width=0.5\textwidth]{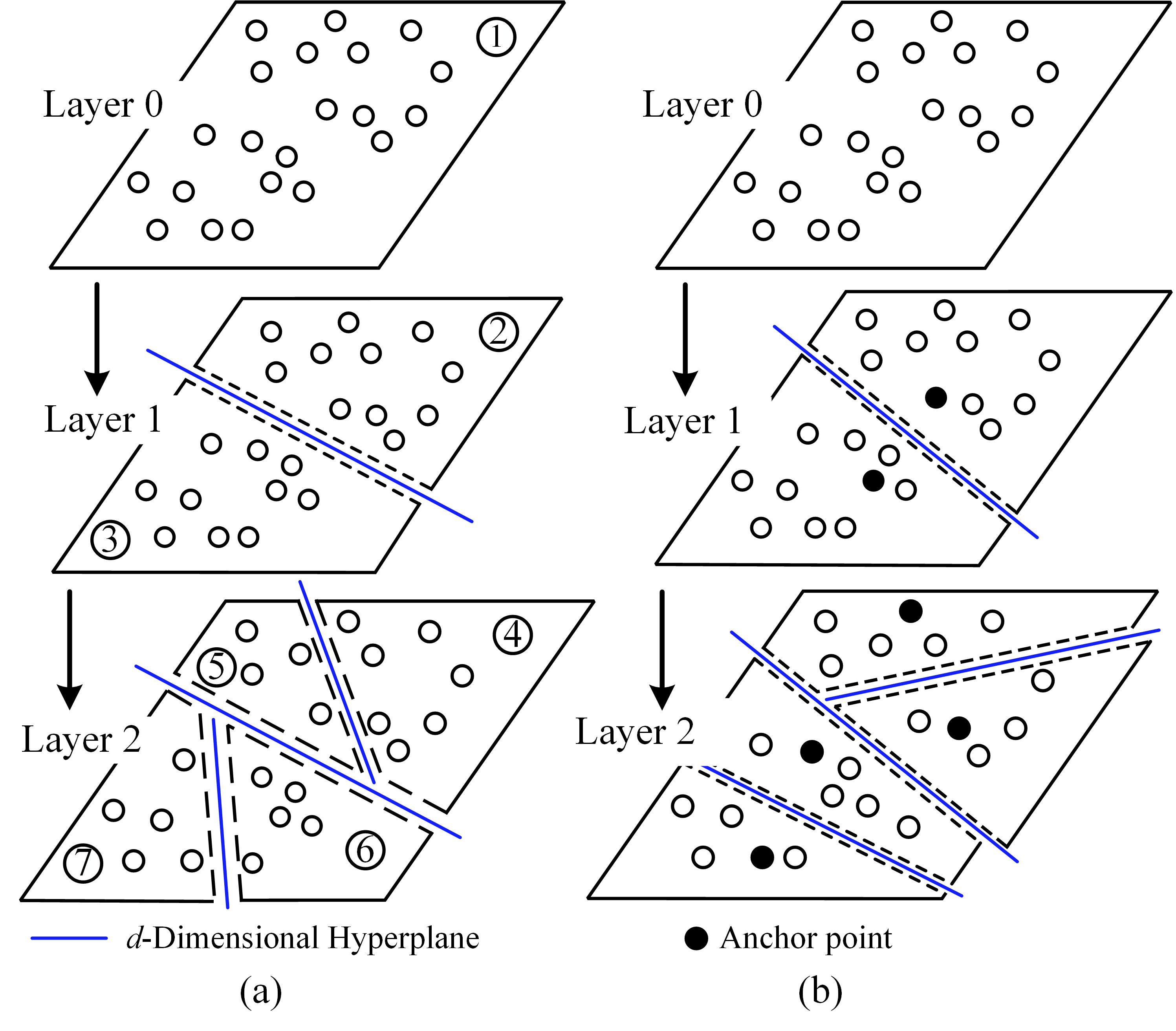}
    \caption{Comparison of the partitioning mechanisms in traditional tree-based ANN and Edge-ANN. (a) Traditional tree-based ANN employs explicitly defined and stored $d$-dimensional hyperplanes for recursive partitioning. (b) Edge-ANN implicitly defines partitions using pairs of anchor points selected from the dataset, thereby avoiding the storage of high-dimensional vectors.}
    \label{fig:annoy_vs_edgeannoy_workflow}
\end{figure*}

The storage efficiency of Edge-ANN, illustrated in \autoref{fig:annoy_vs_edgeannoy_workflow}, stems from a fundamental representational shift. Unlike conventional tree-based method, which stores a high-dimensional normal vector per partition, Edge-ANN requires only two anchor IDs and a scalar offset to define a hyperplane. To ensure this efficient representation does not lead to tree imbalance, we introduce the \textbf{“Binary Anchor Optimization”} algorithm. It selects an anchor pair that approximately bisects the node’s samples and fine-tunes the offset to optimize balance, thereby mitigating adverse effects on retrieval cost and maintaining strong performance.

Experimental results on three large-scale remote sensing datasets, MillionAID \cite{long2021creating}, High-resolution Urban Complex Dataset(Hi-UCD) \cite{tian2022large}, and GlobalUrbanNet Dataset (GUN) \cite{zhong2023global}, empirically validate the superiority of Edge-ANN. Under strict storage constraints, it reduces the secondary storage by approximately 30–40\% compared to ANNoy, while incurring only a modest 3–5\% recall drop. Furthermore, under both low primary and secondary storage constraints, Edge-ANN delivers a superior accuracy-latency trade-off compared to state-of-the-art methods, including HNSW, KD-Tree, and Ball Tree, setting a new benchmark for storage-efficient ANN on remote sensing edge devices.\par

The primary contributions of our work are summarized  as follows:
\begin{itemize}
    \item \textbf{A Dimension-Independent Lightweight Architecture:} We propose an implicit hyperplane representation based on an "anchor pair + scalar offset" mechanism. This design reduces the model's space complexity from $O(t_n(dN/T+N))$ to $O(t_n(N/T+N))$, where $N$ is the number of data, $t_n$ is the number of trees, $T$ is the threshold of leaf node size, significantly decreasing the model size and decoupling it from the feature dimension $d$, which is critical for adapting to the primary and secondary storage constraints of edge devices.
    
    \item \textbf{A Balanced Partitioning Mechanism via Binary Anchor Optimization:} We introduce a novel optimization algorithm to improve partition balance during tree construction. By selecting anchor pairs that approximately bisect the data subset and fine-tuning the hyperplane offset, our method mitigates the performance degradation typically induced by tree imbalance. This approach effectively balances retrieval efficiency and recall accuracy.
     
    \item \textbf{Comprehensive Edge-Oriented Validation and a New Performance Benchmark:} Through systematic experiments on the GUN, Hi-UCD, and MillionAID datasets under tight resource budgets, we demonstrate that Edge-ANN achieves significant gains in storage efficiency and inference speed while preserving competitive retrieval accuracy. This study establishes a new performance benchmark for resource-constrained remote sensing feature retrieval on edge platforms.
\end{itemize}

\section{Related Works}
\label{sec:related_works}

Image feature extraction is the foundational first step in remote sensing feature retrieval, as it directly determines subsequent retrieval efficacy. The subsequent similarity search within the feature space is the pivotal focus of this paper. Existing methods in this domain can be broadly categorized into three paradigms: graph-based, tree-based, and hybrid graph-tree approaches.

\subsection{Remote Sensing Retrieval}

Early research in image description predominantly relied on hand-crafted features. For instance, Aptoula et al. investigated the use of multi-scale texture features, circular covariance histograms, and rotation-invariant point triplets in remote sensing feature  retrieval, introducing a novel descriptor derived from the Fourier power spectrum that delivered competitive performance even with low-dimensional representations \cite{aptoula2013remote}. Scott et al. proposed an entropy-balanced bitmap tree for shape-based object retrieval, a method capable of precisely capturing object contours and supporting efficient retrieval from large-scale image repositories \cite{scott2010entropy}. Yang et al. combined both salient and dense SIFT features, encoding them via the bag-of-visual-words model into compact retrieval vectors, which considerably improved the effectiveness of global feature-based retrieval systems \cite{yang2012geographic}.\par

In recent years, deep neural networks have shown exceptional performance in extracting discriminative features from remote sensing imagery. Tong et al. systematically evaluated deep learning-based features for remote sensing retrieval, reporting state-of-the-art results on several public benchmarks \cite{tong2019exploiting}. Tang et al. adopted convolutional autoencoders to learn representative features, which were then encoded using a bag-of-visual-words framework to facilitate the retrieval of high-spatial-resolution remote sensing images \cite{tang2018unsupervised}. Similarly, Babenko et al. utilized activations from convolutional layers as local descriptors and aggregated them through feature encoding techniques to form global image representations suitable for retrieval tasks \cite{babenko2014neural}. These deep features excel at capturing semantically rich information and complex visual patterns, thereby providing a solid foundation for accurate similarity measurement and efficient feature retrieval.

\subsection{Graph-Based Retrieval Methods}

Graph-based retrieval methods leverage graph structures to index and search high-dimensional target data. The core principle involves constructing a graph $G(V, E)$, where the vertices $V$ represent image features and the edges $E$ encode the similarity (i.e., the distance in the feature space) between their connected endpoints. Consequently, the remote sensing feature retrieval problem is transformed into a task of finding the shortest path or proximal nodes to a given query image on the graph.

Several prominent algorithms exemplify this approach. Navigable Small World (NSW) \cite{malkov2014approximate} is an ANN search algorithm that builds a network with a balance of precise short-range connections and efficient long-range "jumps," enabling effective searching . Hierarchical Navigable Small World (HNSW) \cite{malkov2018efficient} extends this concept by constructing a hierarchical series of NSW graphs to further boost search efficiency. Similarly, NSG enhances search performance by creating a sparse graph to shorten search path lengths, while KGraph \cite{dong2011efficient} starts with a \textit{k}-NN graph and iteratively optimizes its connectivity to improve performance.

However, graph-based methods suffer from a fundamental drawback: their model structures necessitate storing extensive node and edge information, resulting in substantial storage overhead. This becomes prohibitively expensive for large-scale, high-dimensional data, thereby rendering them unsuitable for the stringent lightweight requirements of remote sensing edge devices.

\subsection{Tree-Based Retrieval Methods}

Tree-based retrieval methods offer an effective strategy for ANN search. In contrast to graph-based approaches, which rely on connectivity, tree-based methods accelerate the search process by hierarchically partitioning the feature space. The core idea is to recursively divide the data into progressively smaller subregions, with each node in the tree storing summary information about its corresponding partition. This structure typically yields a search time complexity of $O(\log N)$.

The KD-Tree \cite{zhou2008real}, a foundational algorithm in this category, employs axis-aligned hyperplanes to partition the space and was originally designed for exact \textit{k}-NN and range searches, primarily in low-dimensional settings. To address the performance degradation of KD-Trees in higher dimensions, the Ball Tree \cite{omohundro1989five} was developed, which utilizes hyperspheres for partitioning and is particularly well-suited for non-uniformly distributed data. More recently, ANNoy was introduced to tackle the rapid approximate retrieval of large-scale, high-dimensional vectors by constructing binary trees using random hyperplanes, enhancing both its ease of use and performance in production environments.

Despite their efficiency, a persistent limitation of these tree-based models is their reliance on storing high-dimensional partitioning elements—such as the normal vectors of hyperplanes—which still consumes substantial storage space.

\subsection{Hybrid Graph-and-Tree-Based Retrieval Models}

Hybrid graph-and-tree retrieval models seek to strike a balance between search efficiency and resource consumption by synergizing the local connectivity of graph structures with the global hierarchy of tree structures \cite{dinh2024using}. These models typically employ a layered architecture: the upper layer utilizes a tree structure (e.g., KD-Tree) to rapidly localize a coarse-grained region within the data distribution, while the lower layer performs a fine-grained neighbor search within this localized area using a graph-based method (e.g., HNSW, NSG).

For instance, Microsoft's SPTAG \cite{chen2018sptag} combines a KD-Tree with a graph-based approach, first pruning the candidate set via the tree and then executing iterative neighbor optimization within the graph structure. Similarly, DiskANN \cite{jayaram2019diskann} leverages a graph to store high-precision neighbor relationships while employing the pruning strategy of the Vamana graph (a tree-like structure) to reduce the model's scale.

While this hybrid approach inherits the advantages of both graph and tree algorithms, it also inherits their respective shortcomings. The parameter-tuning process is considerably complex (requiring coordination of tree depth, graph connectivity, etc.), the resulting model architecture is intricate, and the storage remains large. These factors render such models difficult to deploy on storage-constrained remote sensing edge devices.

\section{Methodology}
\label{sec:methodology}

Addressing the challenge of running retrieval models under the severe primary and secondary storage constraints of edge devices, we proposes \textbf{Edge-ANN}, an efficient ANN framework. Our approach departs from the standard practice of storing explicit hyperplanes at each node, which incurs significant dimension-dependent overhead. Instead, we employ an implicit partitioning strategy that leverages pairs of in-dataset points, termed “anchors,” to construct decision hyperplanes implicitly, thereby achieving substantial storage savings. This architectural transformation fundamentally redefines the central problem of node construction: instead of computing an optimal hyperplane, the task becomes a combinatorial optimization problem—identifying an optimal pair of anchors from a subset of data such that their perpendicular bisector most effectively partitions the current feature space. To address this key challenge, we design the \textbf{Binary Anchor Optimization} algorithm, an efficient and novel heuristic designed to rapidly select high-quality anchor pairs from any given data subset. This section begins with a formal statement of the problem, followed by a detailed explanation of the Binary Anchor Optimization algorithm, and concludes with an overview of the full model construction and search procedures.

\subsection{Problem Formulation}
\label{sec:problem_formulation}

The foundational principle of our partitioning strategy is to identify a pair of anchors that divides the dataset as evenly as possible. In cases of identical partitioning balance, the pair that minimizes the sum of distances from data points to the implicitly defined hyperplane is chosen. Consequently, achieving a balanced partition is critical for an efficient tree-based search.

Given a set $S = \{\bm{x}_1, \bm{x}_2, \dots, \bm{x}_N\}$ containing $N$ $d$-dimensional feature vectors, where the cardinality of the set is denoted by $\text{card}(S)$, our objective is to select an optimal pair of distinct anchors $(\bm{p}_i, \bm{p}_j)$ from the set $S$. For any chosen anchor pair, the entire set $S$ is partitioned into two disjoint subsets, $S_i$ and $S_j$, according to the nearest neighbor principle. Formally, each data point $\bm{x} \in S$ is assigned to the subset corresponding to its closer anchor:
\begin{align}
    S_i &= \{\bm{x} \in S \mid \left\lVert\bm{x} - \bm{p}_i\right\rVert_2 \leq \left\lVert\bm{x} - \bm{p}_j\right\rVert_2\} \\
    S_j &= S \setminus S_i
\end{align}

Our goal is to employ a lexicographical optimization approach. We first minimize the size difference between the two subsets, and then, for pairs achieving the same level of balance, we maxmize the sum of distances from the data points to the implicitly defined hyperplane. To this end, we define two objective functions to quantify the quality of an anchor pair:

The primary objective function, $J_1$, measures the imbalance of the partition:
\begin{equation}
    J_1(\bm{p}_i, \bm{p}_j) = \big|\text{card}(S_i) - \text{card}(S_j)\big|
    \label{eq:primary_objective}
\end{equation}

The secondary objective function, $J_2$, measures the sum of distances from the data points to the implicit hyperplane:
\begin{equation}
    J_2(\bm{p}_i, \bm{p}_j) = \sum_{\bm{x} \in S} \frac{\left| \left(\bm{x} - \frac{\bm{p}_i + \bm{p}_j}{2}\right) \cdot (\bm{p}_i - \bm{p}_j) \right|}{\left\lVert\bm{p}_i - \bm{p}_j\right\rVert_2}
    \label{eq:secondary_objective}
\end{equation}

The optimal anchor pair $(\bm{p}_i^*, \bm{p}_j^*)$ is determined through a two-step lexicographical optimization. First, we identify the set of candidate anchor pairs, $P^*$, that minimize the primary objective $J_1$:
\begin{equation}
    P^* = \left\{ (\bm{p}_i, \bm{p}_j) \in S \times S \mid \bm{p}_i \neq \bm{p}_j, J_1(\bm{p}_i, \bm{p}_j) = \min J_1 \right\}
    \label{eq:candidate_set}
\end{equation}
Then, from within $P^*$, we select the anchor pair that maximizes the secondary objective $J_2$:
\begin{equation}
    (\bm{p}_i^*, \bm{p}_j^*) = \underset{(\bm{p}_i, \bm{p}_j) \in P^*}{\text{argmax}} \, J_2(\bm{p}_i, \bm{p}_j)
    \label{eq:final_selection}
\end{equation}

Solving this problem via an exhaustive search requires evaluating all $\binom{N}{2}$ possible anchor pairs, resulting in a computational complexity as high as $O(d \cdot N^2)$. Therefore, designing an efficient heuristic method is imperative.

\subsection{Binary Anchor Optimization Algorithm}
\label{sec:binary_anchor_optimization}

We designed the \textbf{Binary Anchor Optimization} algorithm, an iterative  heuristic approach, to efficiently find a high-quality solution for the optimization problem defined in \autoref{eq:candidate_set} and \autoref{eq:final_selection}. The algorithm is composed of two sequential phases: an iterative search for the optimal anchor pair, followed by a final fine-tuning of the hyperplane's position. The complete process is detailed in \autoref{alg:bipartition2}.

\begin{algorithm}[t]
	\caption{Binary Anchor Optimization}
	\label{alg:bipartition2}
	\begin{algorithmic}[1]
		\Require Set of feature vectors $S$, number of candidates $K$
		\Ensure Optimal anchor points $\bm{p}_1, \bm{p}_2$; Fine-tuning scalar $\Delta d$
		\Statex \textbf{Phase 1: Multi-Candidate Iterative Search}
		\State Initialize $\mathcal{C}$ with $K$ random distinct anchor pairs from $S$
		\State Initialize history of candidate sets $\mathcal{H} \leftarrow \emptyset$
		\While{$\mathcal{C} \notin \mathcal{H}$}
		    \State $\mathcal{H} \leftarrow \mathcal{H} \cup \{\mathcal{C}\}$
		    \State Initialize new candidate set $\mathcal{C}_{\text{new}} \leftarrow \emptyset$
		    \For{each candidate $(\bm{p}_i, \bm{p}_j)$ in $\mathcal{C}$}
		        \State Partition $S$ into $S_i$, $S_j$ by $p_i$, $p_j$
		        \State $(S_{\text{large}}, S_{\text{small}}) \leftarrow \text{card}(S_i) > \text{card}(S_j) ? (S_i, S_j) : (S_j, S_i)$
		        \State $\bm{d}_{\text{diff}} \leftarrow \mathrm{Centroid}(S_{\text{large}}) - \mathrm{Centroid}(S_{\text{small}})$
		        \State $\Delta\bm{v} \leftarrow \alpha \left(\text{card}(S_{\text{small}})/\text{card}(S_{\text{large}})\right) \cdot \bm{d}_{\text{diff}}$
		        \State $\bm{p}_i' \leftarrow \bm{p}_i + \Delta\bm{v}$; \quad $\bm{p}_j' \leftarrow \bm{p}_j + \Delta\bm{v}$
		        \State Partition $S$ into $S_i'$, $S_j'$ by $p_i'$, $p_j'$
		        \State $\bm{p}_i \leftarrow \operatorname*{arg\,min}_{\bm{x} \in S_i'} \left\lVert\bm{x} - \bm{p}_i'\right\rVert_2$
		        \State $\bm{p}_j \leftarrow \operatorname*{arg\,min}_{\bm{x} \in S_j'} \left\lVert\bm{x} - \bm{p}_j'\right\rVert_2$
		        \State $\mathcal{C}_{\text{new}} \leftarrow \mathcal{C}_{\text{new}} \cup \{(\bm{p}_i, \bm{p}_j)\}$
		    \EndFor
		    \State $\mathcal{C} \leftarrow \mathcal{C}_{\text{new}}$
		\EndWhile
		\State $J_1^{\min} \leftarrow \mathcal{C}$
		\State $J_2^{\max} \leftarrow -\infty$
		\For{each pair $(\bm{p}_i, \bm{p}_j)$ in $J_1^{\min}$}
		    \State $J_2 \leftarrow \frac{\sum_{\bm{x} \in S} \left| (\bm{x} - \frac{\bm{p}_i+\bm{p}_j}{2}) \cdot (\bm{p}_j-\bm{p}_i) \right|}{\left\lVert\bm{p}_j-\bm{p}_i\right\rVert_2}$ \Comment{c.f. \autoref{eq:secondary_objective}}
		    \If{$J_2 > J_2^{\max}$}
		        \State $J_2^{\max} \leftarrow J_2$
		        \State $\bm{p}_1 \leftarrow \bm{p}_i$; $\bm{p}_2 \leftarrow \bm{p}_j$
		    \EndIf
		\EndFor
		
		\Statex \textbf{Phase 2: Hyperplane Position Fine-Tuning}
		\State $\bm{w} \leftarrow \bm{p}_2 - \bm{p}_1$
		\State $b_0 \leftarrow (\left\lVert\bm{p}_2\right\rVert_2^2 - \left\lVert\bm{p}_1\right\rVert_2^2) / 2$
		\State $\mathcal{D}_{\text{dist}} \leftarrow \{(\bm{w}^\top \bm{x} - b_0) / \left\lVert\bm{w}\right\rVert_2 \mid \forall \bm{x} \in S\}$ \Comment{c.f. \autoref{eq:signed_distances}}
		\State $\Delta d \leftarrow \mathrm{Median}(\mathcal{D}_{\text{dist}})$
		\State \Return $\bm{p}_1, \bm{p}_2, \Delta d$
	\end{algorithmic}
\end{algorithm}

\subsubsection{Phase 1: Iterative  Anchor Search}
\label{sec:iterative_search}

This phase forms the core of the algorithm, employing a closed-loop "evaluate-adjust-project" iteration to dynamically seek the optimal anchor pair. At the start of each iteration, given a current anchor pair $(\bm{p}_i, \bm{p}_j)$, the dataset $S$ is partitioned into two subsets, $S_i$ and $S_j$. To guide the search toward a more balanced state, we first define an imbalance vector, $\bm{d}_{\text{diff}} = \mathrm{Centroid}(S_{\text{large}}) - \mathrm{Centroid}(S_{\text{small}})$, to quantify the current partition's imbalance, where $S_{\text{small}}$ and $S_{\text{large}}$ are the smaller and larger of the two subsets in terms of cardinality, respectively. Geometrically, this vector points from the centroid of the smaller subset (in terms of cardinality) to the centroid of the larger one, thus providing a directional cue to mitigate the partition imbalance. Our key insight is that translating the separating hyperplane (i.e., the perpendicular bisector of $\bm{p}_i$ and $\bm{p}_j$) in the direction of this imbalance will produce a more balanced partition. We achieve this by generating a pair of "virtual anchors," $(\bm{p}_i', \bm{p}_j')$, which represent idealized anchor positions after this translation: $\bm{p}_i' = \bm{p}_i + \Delta\bm{v}$ and $\bm{p}_j' = \bm{p}_j + \Delta\bm{v}$. The translation vector is defined as $\Delta\bm{v} = \alpha(|S_{\text{small}}|/|S_{\text{large}}|) \cdot \bm{d}_{\text{diff}}$, where the function is defined as $\alpha(\text{ratio})=e^{-2\cdot\text{ratio}^2}$. This exponential function ensures that the translation magnitude is significant for highly imbalanced partitions (when `ratio` is small) and gracefully diminishes as `ratio` approaches 1 (a balanced partition), which facilitates the stable convergence of the algorithm.

However, these virtual anchors are not guaranteed to be members of the original dataset $S$. To adhere to our principle of using only real data points, this idealized solution must be projected back onto $S$. We partition the dataset $S$ again based on the virtual anchors $(\bm{p}_i', \bm{p}_j')$ to obtain temporary subsets $S_i'$ and $S_j'$. Then, the new anchors are chosen as the real data points within each respective subset that are closest to the virtual anchors and are updated to $(\bm{p}_i, \bm{p}_j)$. This projection operation ensures that hyperplanes are always defined by actual data points from the dataset, which is crucial for achieving storage efficiency (as only their IDs need to be stored). While this heuristic approach does not guarantee global optimality, its convergence is assured. Since the dataset $S$ is finite, the number of possible anchor pairs is also finite. Therefore, the algorithm is guaranteed to converge to a local optimum or a previously seen state within a finite number of steps.

\subsubsection{Phase 2: Hyperplane Position Fine-Tuning}
\label{sec:hyperplane_fine-tuning}

The iterative search phase determines the anchor points that define the separating hyperplane. However, due to the discrete nature of real-world data, the standard perpendicular bisector of an anchor pair may not perfectly bisect the dataset. This final phase aims to precisely calibrate the position of the hyperplane.

We define the normal vector of the hyperplane as $\bm{w} = \bm{p}_2 - \bm{p}_1$. Subsequently, we compute the set of signed orthogonal distances from each data point $\bm{x} \in S$ to the initial perpendicular bisector. This set, $\mathcal{D}_{\text{dist}}$, is given by:
\begin{equation}
    \mathcal{D}_{\text{dist}} = \left\{ \frac{\bm{w}^\top \bm{x} - (\left\lVert\bm{p}_2\right\rVert_2^2 - \left\lVert\bm{p}_1\right\rVert_2^2) / 2}{\left\lVert\bm{w}\right\rVert_2} \mid \forall \bm{x} \in S \right\}
    \label{eq:signed_distances}
\end{equation}
To achieve a balanced data partition, the hyperplane must be translated by a distance that causes it to pass precisely through the median of all these projected distances. Therefore, we compute the median of all signed distances: $\Delta d = \mathrm{Median}(\mathcal{D}_{\text{dist}})$. The scalar value $\Delta d$ represents the optimal distance that the hyperplane must be translated along its normal direction. For more details about hyperplane, please refer to \autoref{appendix:hyperplane_derivation}.

\subsection{Model Construction}
\label{sec:model_construction}

Edge-ANN constructs its model by recursively applying the Binary Anchor Optimization algorithm to partition the data, as detailed in \autoref{alg:build}. Starting with the entire dataset, the `BuildNode` procedure is invoked. It first checks if the cardinality of the current subset $S_{\text{current}}$ is less than or equal to the leaf node size threshold $T$. If so, a leaf node containing $S_{\text{current}}$ is created. Otherwise, the Binary Anchor Optimization algorithm (\autoref{alg:bipartition2}) is called on $S_{\text{current}}$ to find the optimal anchor pair $(\bm{p}_1, \bm{p}_2)$ and the fine-tuning scalar $\Delta d$. Subsequently, the dataset is partitioned into $S_1$ and $S_2$ using the derived normal vector $\bm{w}$ and offset $b$. This process is recursively applied to each subset until its size falls below the predefined threshold $T$.

\begin{algorithm}[H]
\caption{Edge-ANN Model Construction}
\label{alg:build}
\begin{algorithmic}[1]
\Require Set of feature vectors $S$, Leaf size threshold $T$
\Ensure Root node of the Edge-ANN tree
\Procedure{BuildNode}{$S_{\text{current}}$}
    \If{$\text{card}(S_{\text{current}}) \le T$}
        \State \Return CreateLeafNode($S_{\text{current}}$)
    \EndIf
    \State $(\bm{p}_1, \bm{p}_2, \Delta d) \leftarrow \text{BinaryAnchorOptimization}(S_{\text{current}})$
    \State $\bm{w} \leftarrow \bm{p}_2 - \bm{p}_1$
    \State $b \leftarrow (\left\lVert\bm{p}_2\right\rVert_2^2 - \left\lVert\bm{p}_1\right\rVert_2^2)/2 + \Delta d \cdot \left\lVert\bm{w}\right\rVert_2$ \Comment{c.f. \autoref{eq:final_offset}}
    \State \Comment{The offset $b$ is adjusted by $\Delta d$ for fine-tuning.}
    \State $S_{1} \leftarrow \{\bm{x} \in S_{\text{current}} \mid \bm{w}^\top \bm{x} - b \le 0\}$
    \State $S_{2} \leftarrow S_{\text{current}} \setminus S_{1}$
    \State $node \leftarrow$ CreateInternalNode(ID of $\bm{p}_1$, ID of $\bm{p}_2$, $\Delta d$)
    \State $node$.left\_child $\leftarrow$ \Call{BuildNode}{$S_{1}$}
    \State $node$.right\_child $\leftarrow$ \Call{BuildNode}{$S_{2}$}
    \State \Return $node$
\EndProcedure
\end{algorithmic}
\end{algorithm}

During the model construction process, particularly when dealing with very large data subsets (especially in the upper layers of the tree), to accelerate the anchor selection, the "Binary Anchor Optimization" step can be applied only to a random sample of $S_{\text{current}}$ (e.g., 10\%). This typically does not significantly compromise the partitioning quality.

\subsection{Retrieval Method}
\label{sec:retrieval_method}

For a query vector, the search process in Edge-ANN is identical to the standard ANNoy algorithm. It involves a top-down traversal of the tree, utilizing a priority queue to manage promising branches for further exploration. At each internal node, the decision of which child node to explore next is determined by the query vector's relative position to the separating hyperplane. The search process terminates after a preset number of leaf nodes have been examined, and the collected candidates are then ranked to produce the final approximate nearest neighbor results.

\section{Experiments and Results}
\label{sec:experiments}

This section aims to systematically evaluate the performance of our proposed Edge-ANN framework through a series of comprehensive experiments. We will first introduce the benchmark datasets, feature extraction methods, and experimental environment configurations employed in our study.

\subsection{Datasets}
\label{sec:datasets}

To evaluate the effectiveness and generalization capabilities of our proposed algorithm, we selected three representative remote sensing datasets for our experiments: MillionAID, Hi-UCD and GUN. These datasets possess distinct characteristics in terms of data sources, scene categories, spatial resolution, and data scale, providing a solid foundation for verifying the algorithm's performance under diverse conditions shown in \autoref{tab:datasets}.

\begin{table*}[t!]
    \centering
    \caption{Remote Sensing Datasets Used for Testing in This Paper}
    \label{tab:datasets}
    \begin{tabularx}{\textwidth}{l >{\raggedright\arraybackslash}X cc}
        \toprule
        \textbf{Dataset} & \textbf{Characteristics} & \textbf{Image Size} & \textbf{Number of Images} \\
        \midrule
        MillionAID & Categories cover natural landscapes, urban infrastructure, agricultural areas, etc. It exhibits extreme diversity in spatial distribution, imaging conditions, and fine-grained categorization. & 256$\times$256, 512$\times$512 & 1 000 000+ \\
        Hi-UCD & This dataset contains rich urban land cover categories: roads, buildings, vegetation, water bodies, etc., and images cover various seasons and lighting conditions, showing significant attribute variations. & 512$\times$512 & 40 800 \\
        GUN & Covers multi-domain, multi-category natural images. Its images are from diverse sources, with fine-grained and well-balanced category labels, suitable for evaluating retrieval accuracy across different scenarios and semantic categories. & Diverse & 1 000 000+ \\
        \bottomrule
    \end{tabularx}
\end{table*}

In this paper, we cropped each image in the Hi-UCD dataset into four 256$\times$256 sub-images for retrieval research, to increase the data volume and unify the scale. To obtain high-quality image content representations, we employed the advanced geospatial foundation model Prithvi-100M \cite{jakubik2023foundation} to extract 768-dimensional deep features for images across all the aforementioned datasets. This model has been extensively pre-trained on massive multispectral remote sensing data, enabling it to capture rich semantic and spatial information, thereby providing high-quality input for subsequent similarity retrieval. In all experiments, we uniformly adopted Euclidean distance as the metric for vectorial similarity in the feature space.

\subsection{Experimental Environment and Constraint Simulation}
\label{sec:experimental_setup}

All experiments were conducted on a server equipped with an AMD EPYC 7K62 48-core CPU and 64GB of primary storage. To precisely simulate the primary storage-constrained operating environment of edge devices, we utilized Docker container technology. By setting the Docker container's memory limit and swap space, we were able to impose different levels of strict primary storage constraints on the runtime environment of each algorithm, thereby reliably evaluating its performance under stroage-constrained conditions.\par

\subsection{Evaluation Metrics}
\label{sec:evaluation_metrics}

The algorithm’s performance was benchmarked along three dimensions: retrieval accuracy, efficiency, and model size. From each dataset, 1000 feature vectors were randomly selected to create a held-out query set, which was excluded from all training phases and utilized for evaluation across all experiments.\par

We employ \textit{Recall} as the core metric to measure retrieval accuracy. Following standard practice, for each query, the algorithm's task is to retrieve the $k$ most similar nearest neighbors \cite{malkov2018efficient}. In this study, we set $k=10$. For a single query, its recall rate is defined as the ratio of the number of nearest neighbors in \textit{Retrieved} that belong to \textit{Ground Truth} to $k$. The final \textit{Recall} is the arithmetic mean of the recall rates for all queries, reflecting the algorithm's overall ability to find relevant results. Its calculation formula is as follows:
\begin{equation}
    \textit{Recall} = \frac{1}{N_q} \sum_{i=1}^{N_q} \frac{\text{card}(\textit{Ground Truth}_i \cap \textit{Retrieved}_i)}{k}
    \label{eq:recall}
\end{equation}
where $N_q$ is the size of the test set, which is 1000 in this paper.

Retrieval efficiency is measured by query time, which directly reflects the algorithm's response speed. We recorded the total time $T_{\text{total}}$ required to execute all $N_q$ queries in the query set, and then calculated its average. This metric's unit is ms/query.
\begin{equation}
    T_{\text{Search}} = T_{\text{total}} / {N_q}
    \label{eq:search_time}
\end{equation}

Model size is evaluated by the secondary storage space (Storage Size) occupied by the model file on disk, with the unit being megabytes (MB). This metric directly relates to the algorithm's deployment feasibility on edge devices with constrained storage and is a key measure of its lightweight nature.

\subsection{Experimental Results}
\label{sec:experimental_results}

We conducted comprehensive experiments on three benchmark datasets: GUN, Hi-UCD, and MillionAID. To systematically evaluate the performance of Edge-ANN, we extensively compared it with the baseline model ANNoy and three widely used ANN algorithms (HNSW, Ball Tree, KD-Tree) across three core dimensions—retrieval accuracy, retrieval efficiency, and model scale—under various primary storage constraints.

\subsubsection{Edge-ANN Versus Baseline}
\label{sec:edge-annoy_vs_annoy}

\begin{figure*}[t!]
    \centering
    \includegraphics[width=0.9\textwidth]{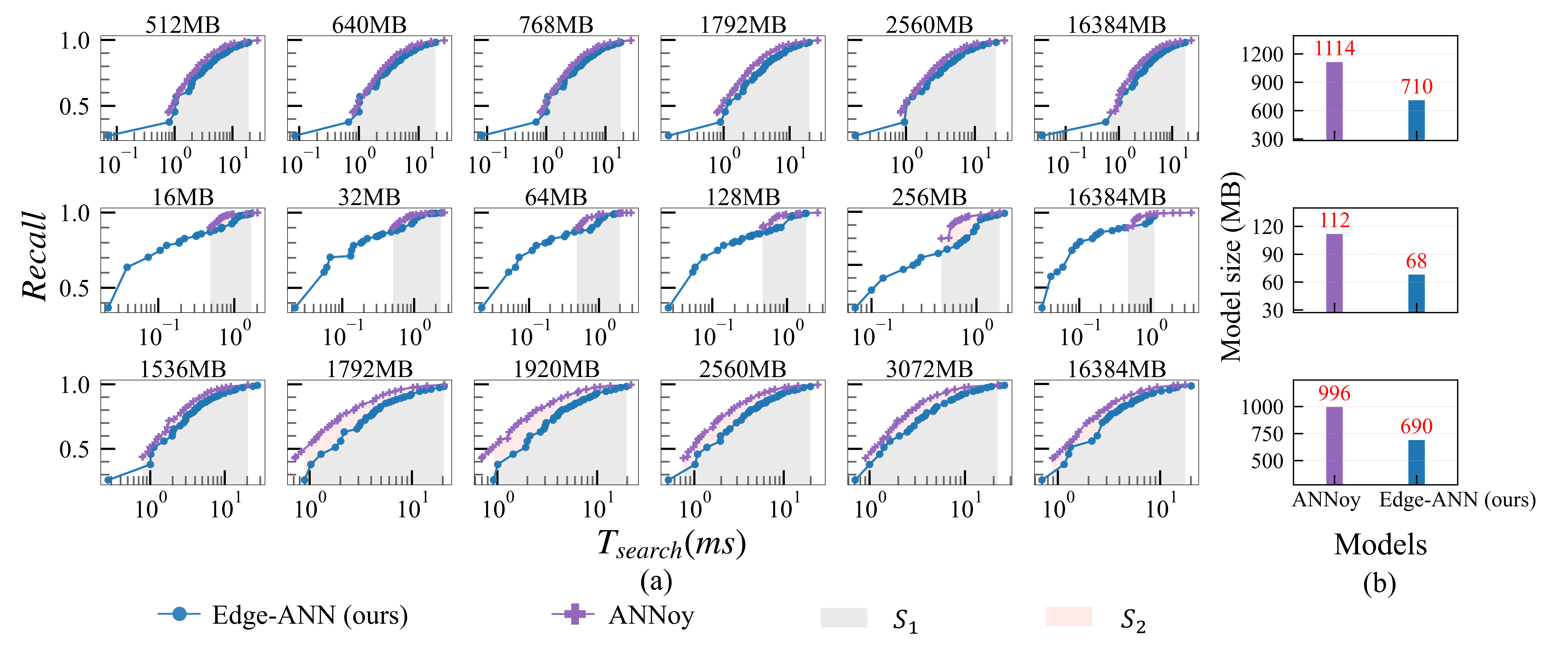}
    \caption{Edge-ANN vs. Baselie Retrieval Performance and Model Size: (a) Results of models under different primary storage constraints across three datasets; (b) Model size, with the first, second, and third rows corresponding to MillionAID, Hi-UCD, and GUN datasets, respectively. For clearer details, please refer to Material 2.}
    \label{fig:edge-annoy_vs_annoy}
\end{figure*}

Given that Edge-ANN is an optimized tree-based method tailored for edge computing, the primary objective of our evaluation is to quantify the trade-off it offers: a substantial reduction in model size at the cost of a minimal, acceptable decrease in retrieval performance. The distinction lies in the method for setting the leaf node splitting threshold: Edge-ANN employs a fixed value of 50, whereas baseline method determines this value adaptively through its built-in algorithm. This difference stems from divergent design philosophies. In Edge-ANN, the threshold of 50 is a pre-defined empirical value, established a priori through a trade-off between index construction efficiency and retrieval accuracy. It serves as a global hyperparameter to achieve a stable and predictable operational balance. \autoref{fig:edge-annoy_vs_annoy} compares the two models directly. It can be observed from \autoref{fig:edge-annoy_vs_annoy}(a) that the retrieval performance of Edge-ANN is marginally lower, representing the trade-off for its model size reduction. This performance gap is quantified as:
\begin{equation}
P_{\text{loss}} = {(S_2 - S_1)}/{S_2}
\label{eq:performance_loss}
\end{equation}
The quantitative results across all datasets are summarized in \autoref{fig:edge-annoy_vs_annoy}. For the MillionAID dataset, the retrieval performance loss ($P_{\text{loss}}$) was -3.82 $\pm$ 0.56\% (percentage $\pm$ SD), while the model size was reduced by 36.24\% (the 1st row in \autoref{fig:edge-annoy_vs_annoy}); for the Hi-UCD dataset, the retrieval performance loss was -3.22 $\pm$ 0.68\%, and the model size was reduced by 39.19\% (the 2nd row in \autoref{fig:edge-annoy_vs_annoy}); for the GUN dataset, the retrieval performance loss was -5.85 $\pm$ 0.96\%, and the model size was reduced by 30.73\% (the 3rd in \autoref{fig:edge-annoy_vs_annoy}). These findings indicate that the proposed method achieves a considerable reduction in model size with only a minor performance sacrifice, highlighting its potential for use on storage-constrained edge devices.

\subsubsection{Edge-ANN Versus Competing Methods}
\label{sec:edge-annoy_vs_competing}

\paragraph{Results on the MillionAID Dataset:}

\begin{figure*}[t!]
    \centering
    \includegraphics[width=0.9\textwidth]{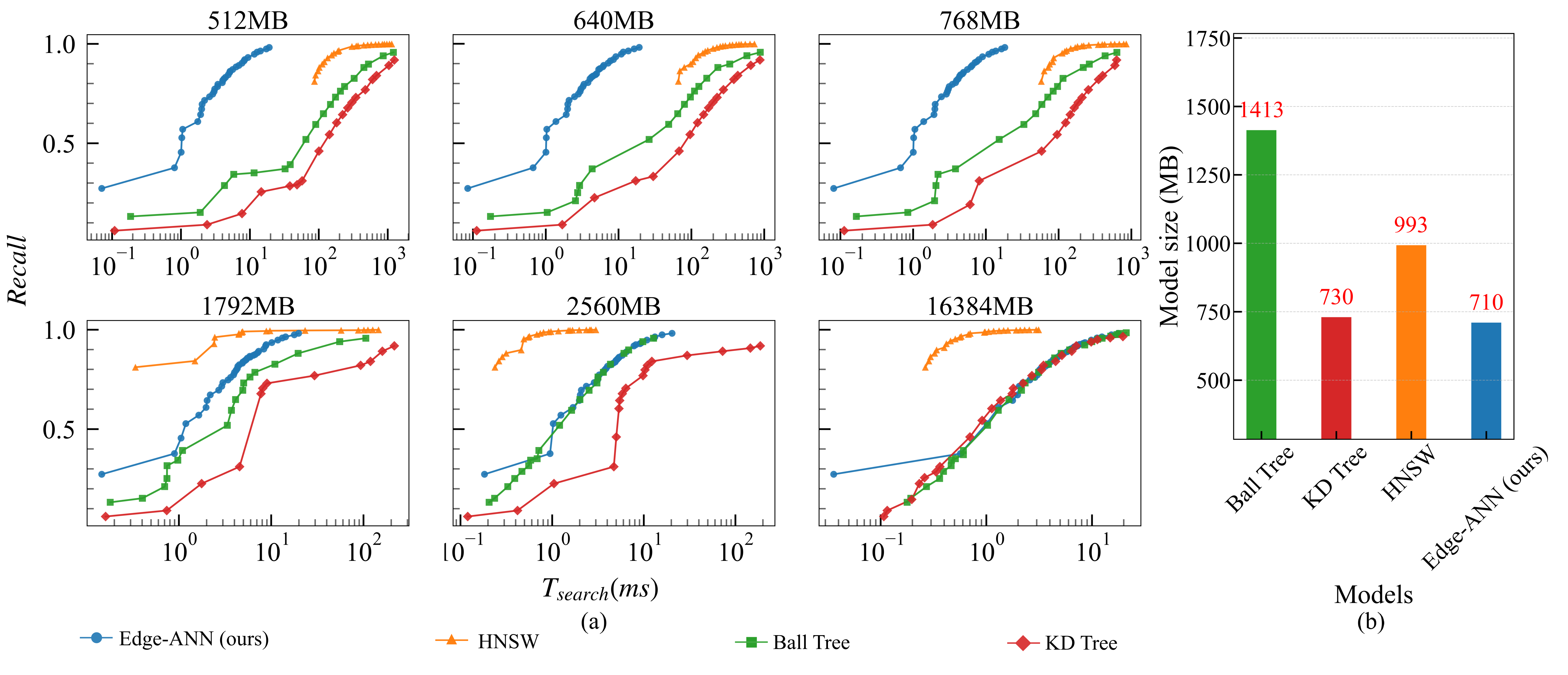}
    \caption{Results on the MillionAID dataset: (a) Retrieval performance of Edge-ANN versus competing methods; (b) Corresponding model sizes.}
    \label{fig:millionaid_results}
\end{figure*}

We simulated six primary storage environments, from severely constrained to abundant: \{512MB, 640MB, 768MB, 1792MB, 2560MB, 16384MB\}. As shown in \autoref{fig:millionaid_results}(a), the experimental results clearly reveal the core advantages of Edge-ANN in primary storage-constrained scenarios. Under lower primary storage configurations (512MB, 640MB, 768MB), Edge-ANN's combined performance in retrieval accuracy and efficiency is significantly superior to all competing methods. As available primary storage increases (1792MB and above), the performance gap between the algorithms gradually narrows, and the performance advantage of HNSW, as the optimal algorithm in resource-rich environments, begins to emerge, which is consistent with its recognized characteristics in such scenarios. At the maximum primary storage of 16384MB, the performance of all tree-based methods tends to converge.\par

Crucially, Edge-ANN maintains the smallest model size across all configurations (\autoref{fig:millionaid_results}(b)). This inherent compactness and its efficiency under low-memory conditions enable a far superior recall rate within a tight latency budget (e.g., $10^0$ ms), maximizing its practical value for edge deployment.

\begin{figure*}[t!]
    \centering
    \includegraphics[width=0.9\textwidth]{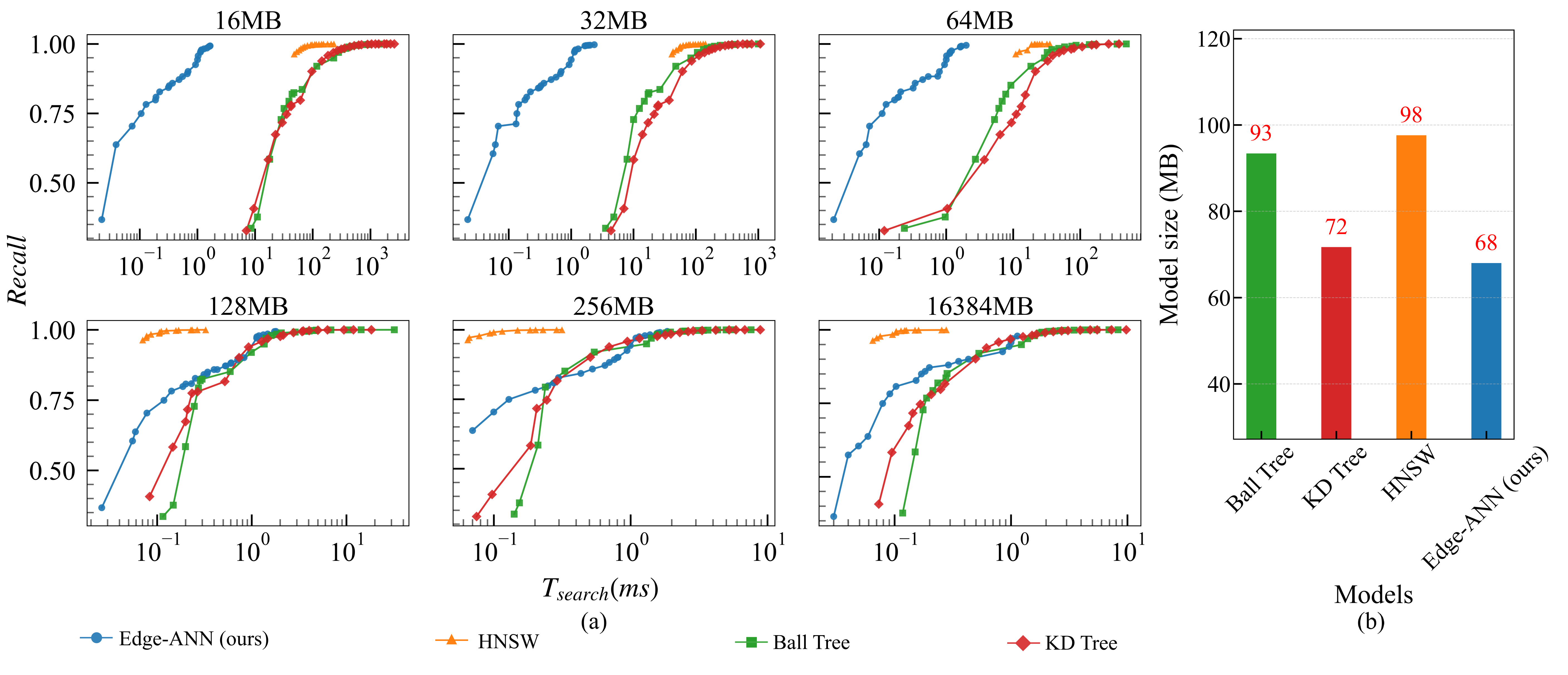}
    \caption{Results on the Hi-UCD dataset: (a) Retrieval performance of Edge-ANN versus competing methods; (b) Corresponding model sizes.}
    \label{fig:hi-ucd_results}
\end{figure*}

\paragraph{Results on the Hi-UCD Dataset:}
Experiments on the Hi-UCD dataset employed a stringent primary storage gradient: \{16MB, 32MB, 64MB, 128MB, 256MB, 16384MB\}. The results in \autoref{fig:hi-ucd_results}(a) show that under extreme constraints (16MB–64MB), Edge-ANN achieves a high recall rate with minimal latency, significantly surpassing Ball Tree, KD-Tree, and HNSW. It also consistently maintains the smallest model size (\autoref{fig:hi-ucd_results}(b)). With more storage available (128MB+), the performance of Edge-ANN and the other tree-based methods becomes comparable, while HNSW demonstrates its strength in resource-abundant conditions. These results further validate Edge-ANN’s robustness and superiority in highly constrained scenarios.

\begin{figure*}[t]
    \centering
    \includegraphics[width=0.9\textwidth]{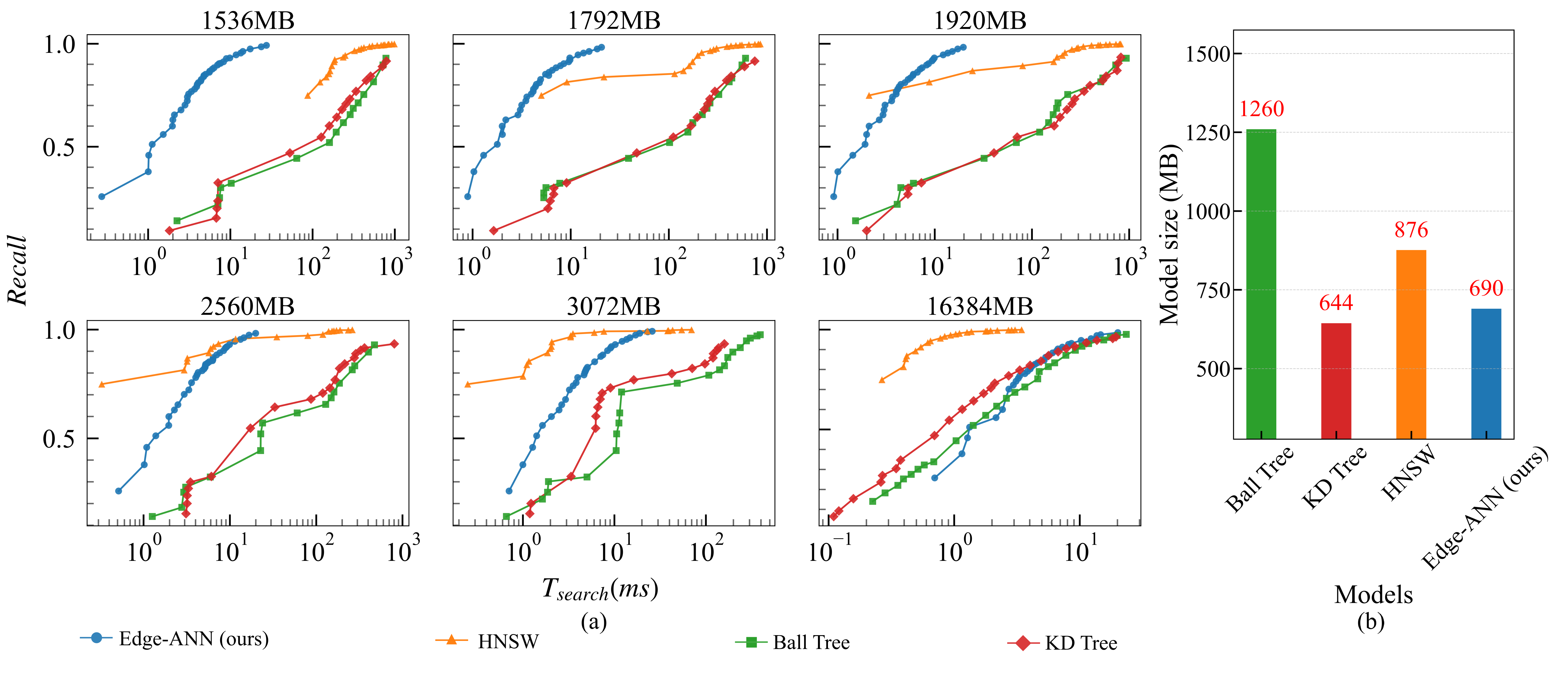}
    \caption{Results on the GUN dataset: (a) Retrieval performance of Edge-ANN versus competing methods; (b) Corresponding model sizes.}
    \label{fig:gun_results}
\end{figure*}

\paragraph{Results on the GUN Dataset:}
In the experiments on the GUN dataset, we selected and configured six primary storage environments: \{1536MB, 1792MB, 1920MB, 2560MB, 3072MB, 16384MB\}. The experimental results (\autoref{fig:gun_results}(a)) once again validate our previous observations. Under low primary storage conditions (1536MB, 1792MB, 1920MB), the performance advantage of Edge-ANN within a given query latency budget is distinctly clear compared to other methods, and it occupies a smaller secondary storage space (\autoref{fig:gun_results}(b)).

In medium primary storage scenarios (2560MB, 3072MB), Edge-ANN's performance begins to be surpassed by HNSW, but it still outperforms Ball Tree and KD-Tree. Finally, under ample primary storage conditions (16384MB), HNSW again exhibits a strong performance advantage, while the performance curves of all tree-based methods gradually converge, showing a consistent level of performance.

\section{Discussion}
\label{sec:discussion}

\subsection{Model Construction Efficiency Analysis}
\label{sec:construction_efficiency}

\begin{figure*}[t!]
    \centering
    \includegraphics[width=0.9\textwidth]{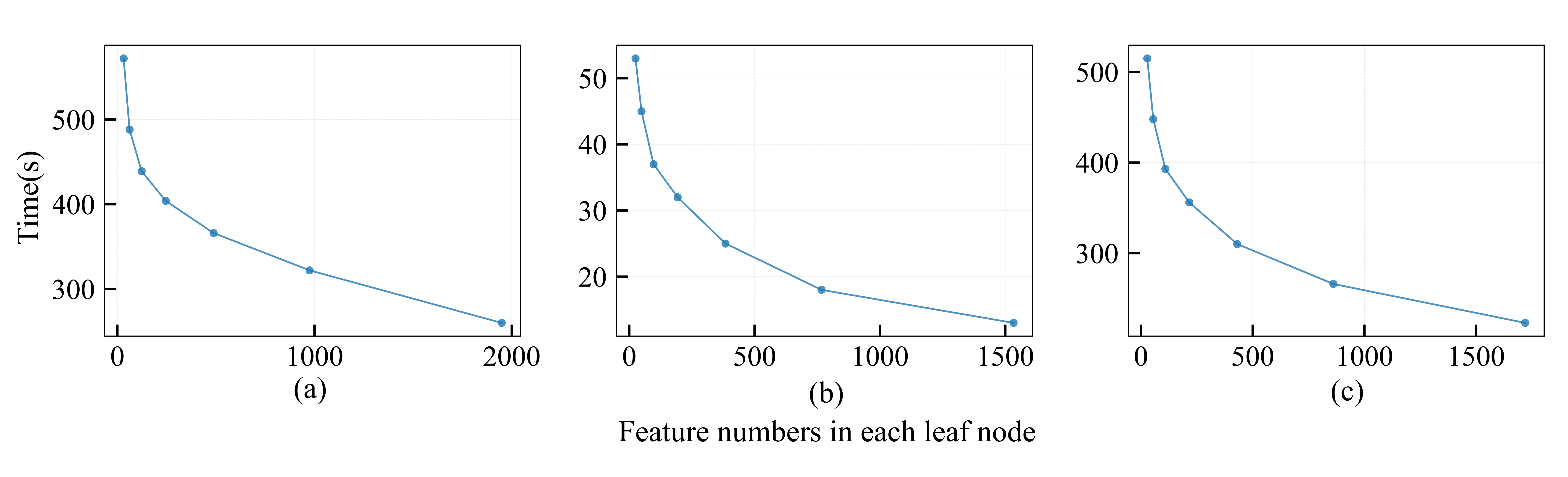}
    \caption{Impact of different leaf node sizes on model construction time: (a) MillionAID; (b) Hi-UCD; (c) GUN.}
    \label{fig:leaf_node_impact}
\end{figure*}

The leaf node size (i.e., the maximum number of features a leaf node can contain) is a critical and adjustable hyperparameter in Edge-ANN, directly governing the trade-off between model construction speed and query performance. The experimental results, as clearly depicted in \autoref{fig:leaf_node_impact}, reveal a significant inverse relationship between the model construction time and the leaf node size threshold. This pattern holds consistently across all three datasets. The fundamental reason is that a larger leaf node size threshold leads to the early termination of the tree's recursive partitioning, thereby constructing a shallower tree. This directly reduces the total number of calls to the recursive partitioning function (i.e., the Binary Anchor Optimization algorithm), thus significantly decreasing the model construction time.

However, this efficiency gain comes at a cost. Larger leaf nodes mean that during the final stage of a query, the number of feature vectors requiring a linear scan within the leaf node also increases, which directly adds to the query latency. Therefore, the selection of the leaf node size is essentially an optimization problem: smaller values favor query speed, while larger values benefit model construction. This inherent trade-off characteristic highlights the flexibility of Edge-ANN, allowing developers to tailor the optimal model structure according to specific application scenarios (e.g., prioritizing rapid deployment versus extreme query response speed).

\begin{figure*}[t!]
    \centering
    \includegraphics[width=0.4\textwidth]{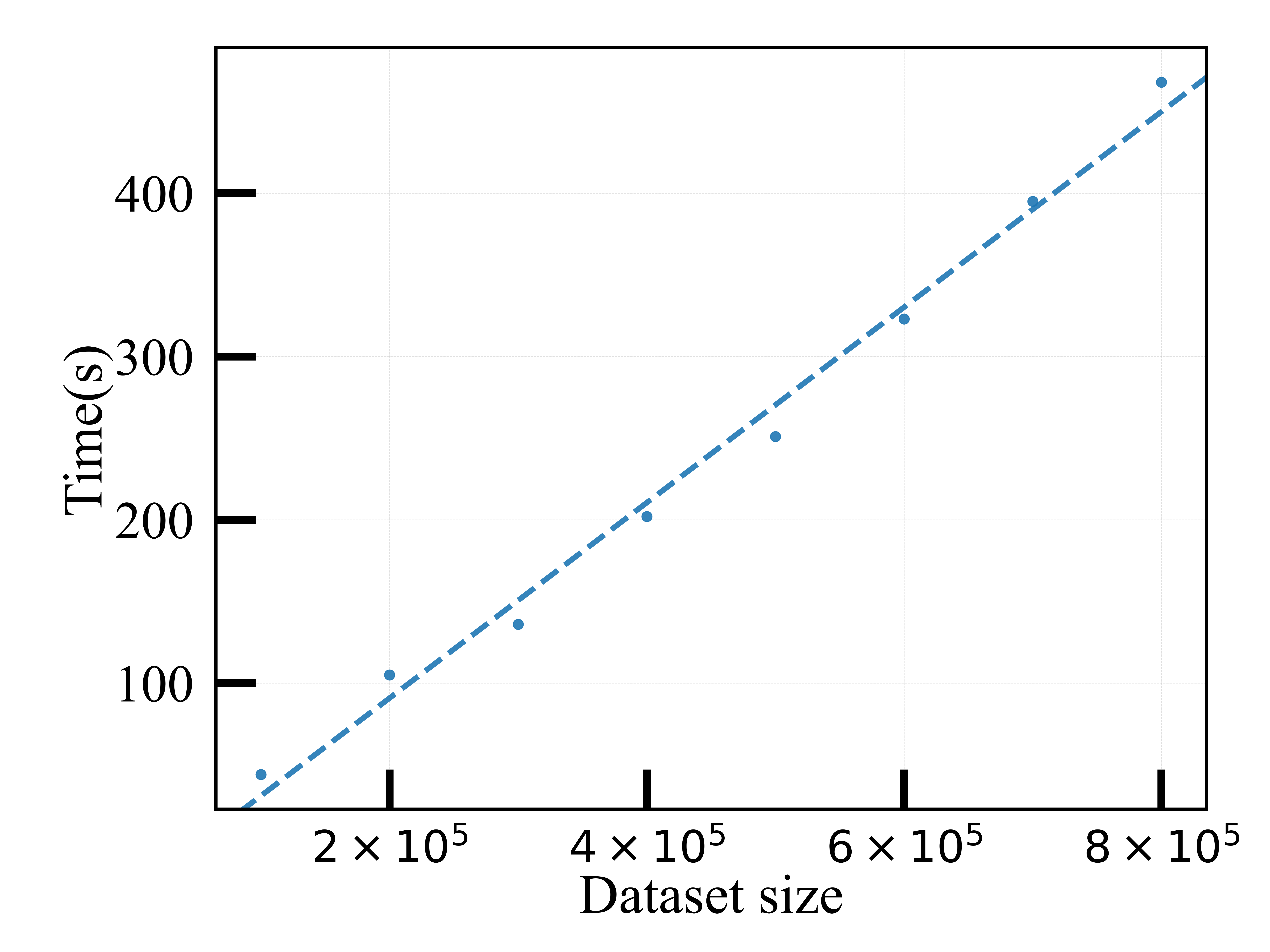}
    \caption{Model construction time of Edge-ANN on GUN sub-datasets.}
    \label{fig:scalability_analysis}
\end{figure*}

Scalability, the performance of an algorithm as it handles growing datasets, is a core criterion for measuring the practical value of a retrieval algorithm. We empirically evaluated the construction time complexity of Edge-ANN by building models on subsets of the GUN dataset of varying scales. For this paper, we constructed a set of subsets from the GUN dataset with sizes of {100k, 200k, 300k, 400k, 500k, 600k, 700k, 800k}. As shown in \autoref{fig:scalability_analysis}, there is a clear, quasi-linear positive correlation between the model construction time and the dataset size \textit{N}. This experimental observation is in complete agreement with our algorithm's theoretical time complexity.

The construction process of Edge-ANN (\autoref{alg:build}) is a recursive tree-building procedure. At each level of the tree, the data of the current node must be partitioned, an operation with $O(N)$ complexity. For a nearly balanced binary tree, its depth is $O(logN)$. Therefore, the overall theoretical time complexity of the construction is $O(NlogN)$. This efficient $O(NlogN)$ complexity demonstrates that Edge-ANN possesses excellent scalability, capable of handling the processing tasks of large-scale remote sensing image datasets.

\subsection{Impact of Leaf Node Size on Retrieval Performance}
\label{sec:leaf_node_performance}

\begin{figure*}[t!]
    \centering
    \includegraphics[width=0.9\textwidth]{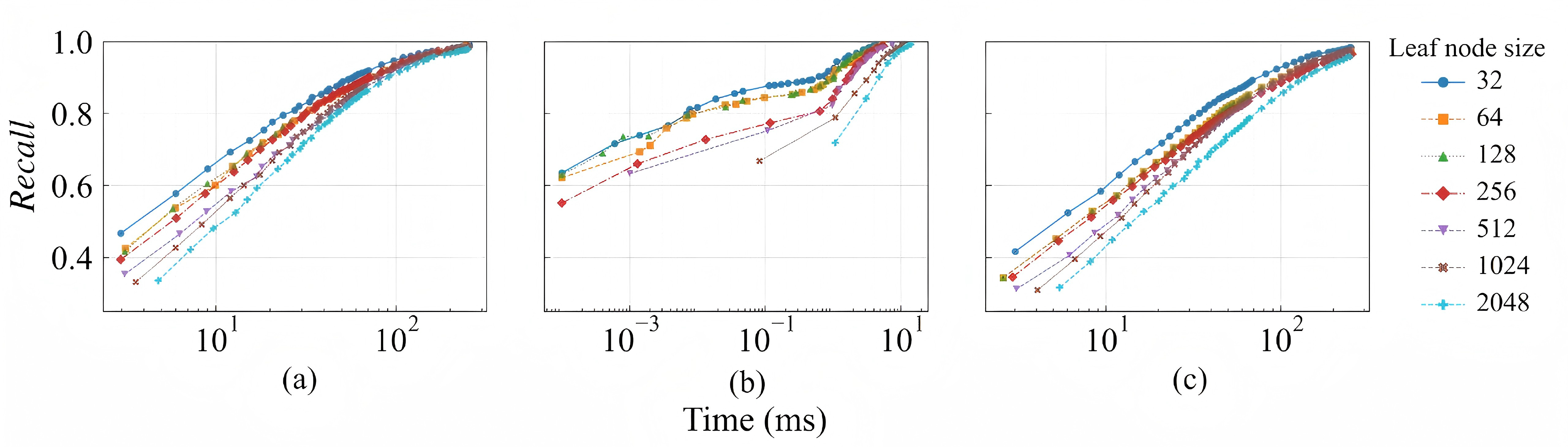}
    \caption{Impact of different leaf node sizes on retrieval performance: (a) Test results on MillionAID, (b) Test results on Hi-UCD, (c) Test results on GUN.}
    \label{fig:leaf_node_performance_impact}
\end{figure*}

The leaf node size not only affects model construction but is also the core hyperparameter for regulating the query performance of Edge-ANN. To investigate its specific impact, we evaluated the algorithm's \textit{Tsearch-Recall} performance curve across a series of leaf node sizes: \{32, 64, 128, 256, 512, 1024\}. As shown in \autoref{fig:leaf_node_performance_impact}, the experimental results reveal a clear and consistent pattern: a smaller leaf node size yields a better \textit{Tsearch-Recall} performance curve. Specifically, for any given latency budget, a model using a smaller leaf node size consistently achieves a higher recall rate. The underlying reason is that the final step of the query process involves a linear scan within the selected leaf nodes. A smaller leaf node size means fewer candidate features need to be processed in this stage. Consequently, within the same total time, more time can be allocated to traversing the tree structure and exploring more potential branches, ultimately improving retrieval accuracy.

Further observation shows that when a sufficiently long query time is allowed, the performance curves for different leaf node sizes gradually converge. This indicates that in applications insensitive to query latency, the impact of the leaf node size on the final retrieval accuracy diminishes. However, in scenarios with stringent real-time requirements, such as edge computing, a smaller leaf node size is crucial for achieving low-latency, high-accuracy retrieval.

Synthesizing the analysis from \autoref{sec:construction_efficiency}, we uncover a fundamental trade-off introduced by this hyperparameter:
\begin{itemize}
    \item \textbf{Decreasing the leaf node size:} Significantly improves query performance (a better \textit{Tsearch-Recall} curve) at the cost of a substantial increase in model construction time.
    \item \textbf{Increasing the leaf node size:} Significantly accelerates model construction speed but sacrifices some query latency.
\end{itemize}

Therefore, in practical applications, the choice of this parameter should be based on a comprehensive consideration of specific requirements. For instance, a dynamic dataset that requires frequent model updates might necessitate a larger value to ensure construction efficiency, whereas a static dataset built once for long-term use should opt for a smaller value to maximize query performance. Based on our experience, we recommend avoiding setting this value below 32 to prevent the model construction time from becoming excessively long.

\subsection{The Primary Storage Advantage Zone of Edge-ANN}
\label{sec:advantage_zone}

We have already demonstrated that HNSW is the top-performing algorithm when resources are abundant. However, this conclusion no longer holds in storage-constrained edge environments. To provide clear guidance for practical applications, this section aims to precisely define the performance advantage boundary of Edge-ANN relative to HNSW. We conducted a series of experiments based on subsets of the GUN dataset of varying scales, systematically investigating how the two key variables of available primary storage and dataset size jointly affect their relative performance. The core objective is to create a "Performance Phase Diagram" to clearly indicate which algorithm should be prioritized under different resource configurations.

\begin{figure*}[t!]
    \centering
    \includegraphics[width=0.8\textwidth]{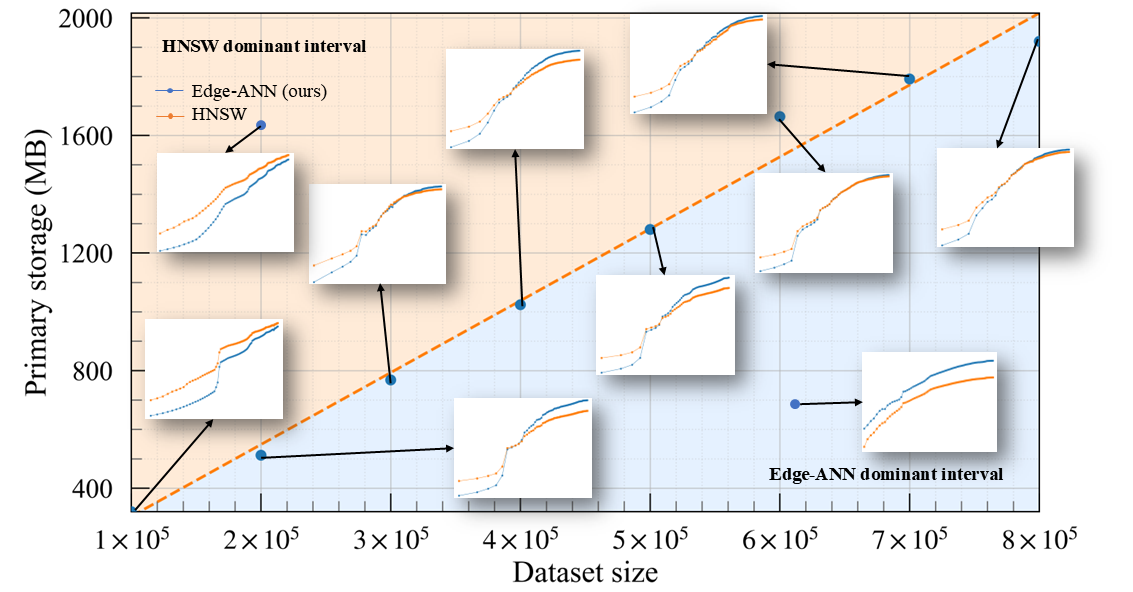}
    \caption{Performance Phase Diagram of Edge-ANN and HNSW.}
    \label{fig:phase_diagram}
\end{figure*}

The experimental results, as shown in \autoref{fig:phase_diagram}, reveal a key pattern: the "critical point" of performance between Edge-ANN and HNSW—where their performances are comparable—exhibits a clear linear relationship in the "primary storage -- dataset size" coordinate system. This linear boundary effectively divides the application scenarios into two distinct advantage zones:
\begin{itemize}
    \item \textbf{Edge-ANN Advantage Zone (lower right, light blue area):} This region represents scenarios with a low ratio of "primary storage / dataset size", i.e., processing large-scale datasets under limited primary storage conditions. When an application's resource configuration falls into this zone, choosing Edge-ANN will result in a significant performance improvement.
    \item \textbf{HNSW Advantage Zone (upper left):} This region corresponds to scenarios with a high ratio of "primary storage / dataset size", where primary storage resources are relatively abundant. In such cases, selecting HNSW is more appropriate.
\end{itemize}

Therefore, this linear boundary provides a powerful and intuitive decision heuristic for algorithm selection. Developers can locate their position on this "phase diagram" based on the actual primary storage budget of their edge device and the size of the dataset to be processed, allowing them to scientifically and efficiently choose the most suitable retrieval algorithm for their specific application scenario.

\subsection{Algorithmic Complexity of Edge-ANN}
\label{sec:algorithmic_complexity}

In the original ANNoy algorithm, each non-leaf node of a tree needs to store a hyperplane for partitioning the space, which is defined by a normal vector of the same dimension $d$, as the data. Consequently, the space overhead of its model file is directly related to the data dimension $d$, and the total space complexity can be expressed as $O(t_n(dN/T + N))$.

In contrast, with Edge-ANN's "Binary Anchor Optimization" strategy, each non-leaf node only needs to store the IDs (integers) of two anchors. This reduces the storage overhead per node from $O(d)$ to $O(1)$, thereby completely eliminating the model structure's dependency on the data dimension. Therefore, its space complexity is $O(t_n(N/T + N))$. The retrieval time complexity for Edge-ANN includes the tree traversal time complexity, $O(\log(N/T))$, and the in-leaf scanning time complexity, $O(T)$. Thus, the total retrieval time complexity is $O(\log(N/T) + T)$.\par

\section{Conclusion}
\label{sec:conclusion}

Addressing the challenge of efficient retrieval for remote sensing images on edge devices with constrained primary and secondary storage, this paper introduces a lightweight ANN algorithm, Edge-ANN. The core innovation of this method lies in the introduction of a binary anchor optimization strategy to replace the high-dimensional hyperplanes used in traditional methods, thereby achieving a balanced partitioning of the dataset. This key design completely eliminates the dependency of the model's storage on feature dimensions, shifting the model's space complexity from being dimension-related to being solely dependent on the number of features, ultimately leading to a significant reduction in model size.

Comprehensive experimental evaluations demonstrate that under low primary storage conditions, Edge-ANN's retrieval efficiency and accuracy are significantly superior to current mainstream methods. Concurrently, we also confirmed the dominant performance of HNSW when storage resources are abundant. To this end, we further experimentally mapped out the performance advantage zone of Edge-ANN relative to HNSW, providing an important, quantifiable reference for algorithm selection in edge computing scenarios. The current work validates the excellent performance of this method in Euclidean distance space. Future research will focus on extending and optimizing it for other key distance metric spaces, such as Manhattan distance and cosine distance, to further broaden its applicability across different remote sensing application scenarios.

\section*{CRediT authorship contribution statement}
Xianwei Lv: Conceptualization, Methodology, Writing, Funding acquisition. Debin Tang: Experiments, Coding, Writing, review \& editing. Zhecheng Shi: Coding, Data Curation.  Wang Wang: Methodology, Writing.  Yujiao Zheng: Data Curation, Investigation. Xiatian Zhu: Review \& Supervision.

\section*{Acknowledgement}
\noindent This work was supported by Xianwei's grants from the Scientific Research Startup Fund of Northeastern University at Qinhuangdao (9060212362301), the Fundamental Research Funds for the Central Universities (N2423007), and Youth Program of the Hebei Provincial Natural Science Foundation (D2025501005).

\section*{Declaration of generative AI and AI-assisted technologies in the 
writing process}
\noindent During the preparation of this work the authors used Gemini2.5 pro in 
order to improve language. After using this tool, the authors reviewed 
and edited the content as needed and take full responsibility for the 
content of the publication.

\section*{Declaration of competing interest}
\noindent We declare that we do not have any commercial or associative interest that represents a conflict of interest in connection with the work submitted.

\section*{Data availability}
\noindent Data will be made available on request.

\bibliographystyle{elsarticle-num-names}

\bibliography{rmfile} 

@article{deren2014automatic,
  title={Automatic analysis and mining of remote sensing big data},
  author={Deren, LI and Liangpei, Zhang and Guisong, Xia},
  journal={Acta Geodaetica et Cartographica Sinica},
  volume={43},
  number={12},
  pages={1211},
  year={2014}
}

@article{shi2025satellite,
  title={Satellite edge artificial intelligence with large models: architectures and technologies},
  author={Shi, Yuanming and Zhu, Jingyang and Jiang, Chunxiao and Kuang, Linling and Letaief, Khaled Ben},
  journal={Science China Information Sciences},
  volume={68},
  number={7},
  pages={170302},
  year={2025},
  publisher={Springer}
}

@article{khankeshizadeh2024novel,
  title={A novel weighted ensemble transferred U-Net based model (WETUM) for postearthquake building damage assessment from UAV data: A comparison of deep learning-and machine learning-based approaches},
  author={Khankeshizadeh, Ehsan and Mohammadzadeh, Ali and Arefi, Hossein and Mohsenifar, Amin and Pirasteh, Saied and Fan, En and Li, Huxiong and Li, Jonathan},
  journal={IEEE Transactions on Geoscience and Remote Sensing},
  volume={62},
  pages={1--17},
  year={2024},
  publisher={IEEE}
}

@article{zhao2019review,
  title={Review of scene matching visual navigation for unmanned aerial vehicles},
  author={Zhao, Chunhui and Zhou, Yihui and Lin, Zhao and Hu, J and Pan, Q},
  journal={Scientia Sinica Informationis},
  volume={49},
  number={5},
  pages={507--519},
  year={2019}
}

@article{li2022development,
  title={Development and Application of a Smart Emergency Response Platfrom for Earthquake Disasters Based on Multi-Source Monitoring Data},
  author={Li, W and Wang, Q and Cheng, W and Xie, X and Du, J and Yan, H},
  journal={The International Archives of the Photogrammetry, Remote Sensing and Spatial Information Sciences},
  volume={48},
  pages={25--30},
  year={2022},
  publisher={Copernicus GmbH}
}

@article{zhou2017predicting,
  title={Predicting grain yield in rice using multi-temporal vegetation indices from UAV-based multispectral and digital imagery},
  author={Zhou, Xiang and Zheng, HB and Xu, XQ and He, JY and Ge, XK and Yao, Xia and Cheng, Tao and Zhu, Yan and Cao, WX and Tian, YC},
  journal={ISPRS Journal of Photogrammetry and Remote Sensing},
  volume={130},
  pages={246--255},
  year={2017},
  publisher={Elsevier}
}

@article{zhang2022expandable,
  title={Expandable on-board real-time edge computing architecture for Luojia3 intelligent remote sensing satellite},
  author={Zhang, Zhiqi and Qu, Zhuo and Liu, Siyuan and Li, Dehua and Cao, Jinshan and Xie, Guangqi},
  journal={Remote Sensing},
  volume={14},
  number={15},
  pages={3596},
  year={2022},
  publisher={MDPI}
}

@article{garcia2024advancements,
  title={Advancements in onboard processing of synthetic aperture radar (sar) data: Enhancing efficiency and real-time capabilities},
  author={Garc{\'\i}a, Laura Parra and Furano, Gianluca and Ghiglione, Max and Zancan, Valentina and Imbembo, Ernesto and Ilioudis, Christos and Clemente, Carmine and Trucco, Paolo},
  journal={IEEE Journal of Selected Topics in Applied Earth Observations and Remote Sensing},
  volume={17},
  pages={16625--16645},
  year={2024},
  publisher={IEEE}
}

@article{lowe2004distinctive,
  title={Distinctive image features from scale-invariant keypoints},
  author={Lowe, David G},
  journal={International journal of computer vision},
  volume={60},
  number={2},
  pages={91--110},
  year={2004},
  publisher={Springer}
}

@article{santini2002similarity,
  title={Similarity measures},
  author={Santini, Simone and Jain, Ramesh},
  journal={IEEE Transactions on pattern analysis and machine Intelligence},
  volume={21},
  number={9},
  pages={871--883},
  year={2002},
  publisher={IEEE}
}

@article{zhu2024cross,
  title={Cross-modal contrastive learning with spatiotemporal context for correlation-aware multiscale remote sensing image retrieval},
  author={Zhu, Lilu and Wang, Yang and Hu, Yanfeng and Su, Xiaolu and Fu, Kun},
  journal={IEEE Transactions on Geoscience and Remote Sensing},
  volume={62},
  pages={1--21},
  year={2024},
  publisher={IEEE}
}

@article{zhang2024efficient,
  title={Efficient and effective retrieval of dense-sparse hybrid vectors using graph-based approximate nearest neighbor search},
  author={Zhang, Haoyu and Liu, Jun and Zhu, Zhenhua and Zeng, Shulin and Sheng, Maojia and Yang, Tao and Dai, Guohao and Wang, Yu},
  journal={arXiv preprint arXiv:2410.20381},
  year={2024}
}

@article{heo2018distance,
  title={Distance encoded product quantization for approximate k-nearest neighbor search in high-dimensional space},
  author={Heo, Jae-Pil and Lin, Zhe and Yoon, Sung-Eui},
  journal={IEEE transactions on pattern analysis and machine intelligence},
  volume={41},
  number={9},
  pages={2084--2097},
  year={2018},
  publisher={IEEE}
}

@article{esmaeili2012fast,
  title={A fast approximate nearest neighbor search algorithm in the hamming space},
  author={Esmaeili, Mani Malek and Ward, Rabab Kreidieh and Fatourechi, Mehrdad},
  journal={IEEE transactions on pattern analysis and machine intelligence},
  volume={34},
  number={12},
  pages={2481--2488},
  year={2012},
  publisher={IEEE}
}

@article{malkov2018efficient,
  title={Efficient and robust approximate nearest neighbor search using hierarchical navigable small world graphs},
  author={Malkov, Yu A and Yashunin, Dmitry A},
  journal={IEEE transactions on pattern analysis and machine intelligence},
  volume={42},
  number={4},
  pages={824--836},
  year={2018},
  publisher={IEEE}
}

@article{zhou2008real,
  title={Real-time kd-tree construction on graphics hardware},
  author={Zhou, Kun and Hou, Qiming and Wang, Rui and Guo, Baining},
  journal={ACM Transactions on Graphics (TOG)},
  volume={27},
  number={5},
  pages={1--11},
  year={2008},
  publisher={ACM New York, NY, USA}
}

@misc{annoy2015,
  author       = {Bernhardsson, Erik},
  title        = {{Annoy: Approximate Nearest Neighbors in C++/Python}},
  howpublished = {GitHub repository},
  year         = {2015},
  url          = {https://github.com/spotify/annoy},
  note         = {Optimized for memory usage and loading/saving to disk}
}

@article{tian2022large,
  title={Large-scale deep learning based binary and semantic change detection in ultra high resolution remote sensing imagery: From benchmark datasets to urban application},
  author={Tian, Shiqi and Zhong, Yanfei and Zheng, Zhuo and Ma, Ailong and Tan, Xicheng and Zhang, Liangpei},
  journal={ISPRS Journal of Photogrammetry and Remote Sensing},
  volume={193},
  pages={164--186},
  year={2022},
  publisher={Elsevier}
}

@article{zhong2023global,
  title={Global urban high-resolution land-use mapping: From benchmarks to multi-megacity applications},
  author={Zhong, Yanfei and Yan, Bowen and Yi, Jingjun and Yang, Ruiyi and Xu, Mengzi and Su, Yu and Zheng, Zhendong and Zhang, Liangpei},
  journal={Remote Sensing of Environment},
  volume={298},
  pages={113758},
  year={2023},
  publisher={Elsevier}
}

@article{aptoula2013remote,
  title={Remote sensing image retrieval with global morphological texture descriptors},
  author={Aptoula, Erchan},
  journal={IEEE transactions on geoscience and remote sensing},
  volume={52},
  number={5},
  pages={3023--3034},
  year={2013},
  publisher={IEEE}
}

@article{scott2010entropy,
  title={Entropy-balanced bitmap tree for shape-based object retrieval from large-scale satellite imagery databases},
  author={Scott, Grant J and Klaric, Matthew N and Davis, Curt H and Shyu, Chi-Ren},
  journal={IEEE Transactions on Geoscience and Remote Sensing},
  volume={49},
  number={5},
  pages={1603--1616},
  year={2010},
  publisher={IEEE}
}

@article{yang2012geographic,
  title={Geographic image retrieval using local invariant features},
  author={Yang, Yi and Newsam, Shawn},
  journal={IEEE transactions on geoscience and remote sensing},
  volume={51},
  number={2},
  pages={818--832},
  year={2012},
  publisher={IEEE}
}

@article{tong2019exploiting,
  title={Exploiting deep features for remote sensing image retrieval: A systematic investigation},
  author={Tong, Xin-Yi and Xia, Gui-Song and Hu, Fan and Zhong, Yanfei and Datcu, Mihai and Zhang, Liangpei},
  journal={IEEE Transactions on Big Data},
  volume={6},
  number={3},
  pages={507--521},
  year={2019},
  publisher={IEEE}
}

@article{tang2018unsupervised,
  title={Unsupervised deep feature learning for remote sensing image retrieval},
  author={Tang, Xu and Zhang, Xiangrong and Liu, Fang and Jiao, Licheng},
  journal={Remote Sensing},
  volume={10},
  number={8},
  pages={1243},
  year={2018},
  publisher={MDPI}
}

@inproceedings{babenko2014neural,
  title={Neural codes for image retrieval},
  author={Babenko, Artem and Slesarev, Anton and Chigorin, Alexandr and Lempitsky, Victor},
  booktitle={European conference on computer vision},
  pages={584--599},
  year={2014},
  organization={Springer}
}

@article{malkov2014approximate,
  title={Approximate nearest neighbor algorithm based on navigable small world graphs},
  author={Malkov, Yury and Ponomarenko, Alexander and Logvinov, Andrey and Krylov, Vladimir},
  journal={Information Systems},
  volume={45},
  pages={61--68},
  year={2014},
  publisher={Elsevier}
}

@inproceedings{dong2011efficient,
  title={Efficient k-nearest neighbor graph construction for generic similarity measures},
  author={Dong, Wei and Moses, Charikar and Li, Kai},
  booktitle={Proceedings of the 20th international conference on World wide web},
  pages={577--586},
  year={2011}
}

@article{omohundro1989five,
  title={Five balltree construction algorithms},
  author={Omohundro, Stephen M},
  year={1989},
  publisher={International Computer Science Institute Berkeley}
}

@inproceedings{dinh2024using,
  title={Using knowledge graph and KD-tree random forest for image retrieval},
  author={Dinh, Nguyen Thi and Le, Thanh Manh and Van, Thanh The},
  booktitle={World Conference on Information Systems and Technologies},
  pages={13--25},
  year={2024},
  organization={Springer}
}

@misc{chen2018sptag,
  title={SPTAG: A library for fast approximate nearest neighbor search},
  author={Chen, Qi and Wang, Haidong and Li, Mingqin and Ren, Gang and Li, Scarlett and Zhu, Jeffery and Li, Jason and Liu, Chuanjie and Zhang, Lintao and Wang, Jingdong},
  year={2018}
}

@article{jayaram2019diskann,
  title={Diskann: Fast accurate billion-point nearest neighbor search on a single node},
  author={Jayaram Subramanya, Suhas and Devvrit, Fnu and Simhadri, Harsha Vardhan and Krishnawamy, Ravishankar and Kadekodi, Rohan},
  journal={Advances in neural information processing Systems},
  volume={32},
  year={2019}
}

@article{jakubik2023foundation,
  title={Foundation models for generalist geospatial artificial intelligence},
  author={Jakubik, Johannes and Roy, Sujit and Phillips, CE and Fraccaro, Paolo and Godwin, Denys and Zadrozny, Bianca and Szwarcman, Daniela and Gomes, Carlos and Nyirjesy, Gabby and Edwards, Blair and others},
  journal={arXiv preprint arXiv:2310.18660},
  year={2023}
}

@article{long2021creating,
  title={On creating benchmark dataset for aerial image interpretation: Reviews, guidances, and million-aid},
  author={Long, Yang and Xia, Gui-Song and Li, Shengyang and Yang, Wen and Yang, Michael Ying and Zhu, Xiao Xiang and Zhang, Liangpei and Li, Deren},
  journal={IEEE Journal of selected topics in applied earth observations and remote sensing},
  volume={14},
  pages={4205--4230},
  year={2021},
  publisher={IEEE}
}

@article{janakiraman2021study,
  title={Study and Analysis of User Desired Image Retrieval},
  author={Janakiraman, S and others},
  journal={Recent Advances in Computer Science and Communications (Formerly: Recent Patents on Computer Science)},
  volume={14},
  number={8},
  pages={2538--2550},
  year={2021},
  publisher={Bentham Science Publishers}
}

\appendix
\section{Derivation of Hyperplane Equations}
\label{appendix:hyperplane_derivation}
This appendix provides a detailed derivation of the hyperplane equations used in our methodology, ensuring clarity and geometric consistency.

\subsection{Perpendicular Bisector of Two Points}
\label{appendix:perpendicular_bisector}
Given two distinct points, $\bm{p}_1$ and $\bm{p}_2$, in a $d$-dimensional Euclidean space, the perpendicular bisector is the locus of points $\bm{x}$ that are equidistant from $\bm{p}_1$ and $\bm{p}_2$. Mathematically, this is expressed as:
\begin{equation}
    \left\lVert\bm{x} - \bm{p}_1\right\rVert_2^2 = \left\lVert\bm{x} - \bm{p}_2\right\rVert_2^2
\end{equation}
Expanding both sides:
\begin{equation}
    (\bm{x} - \bm{p}_1)^\top(\bm{x} - \bm{p}_1) = (\bm{x} - \bm{p}_2)^\top(\bm{x} - \bm{p}_2)
\end{equation}
\begin{equation}
    \bm{x}^\top\bm{x} - 2\bm{x}^\top\bm{p}_1 + \bm{p}_1^\top\bm{p}_1 = \bm{x}^\top\bm{x} - 2\bm{x}^\top\bm{p}_2 + \bm{p}_2^\top\bm{p}_2
\end{equation}
Subtracting $\bm{x}^\top\bm{x}$ from both sides and rearranging:
\begin{equation}
    2\bm{x}^\top(\bm{p}_2 - \bm{p}_1) = \bm{p}_2^\top\bm{p}_2 - \bm{p}_1^\top\bm{p}_1
\end{equation}
Let $\bm{w} = \bm{p}_2 - \bm{p}_1$. This vector $\bm{w}$ is the normal vector to the hyperplane. The equation becomes:
\begin{equation}
    2\bm{w}^\top\bm{x} = \left\lVert\bm{p}_2\right\rVert_2^2 - \left\lVert\bm{p}_1\right\rVert_2^2
\end{equation}
Dividing by 2, we get the standard form of the hyperplane equation $\bm{w}^\top\bm{x} - b_0 = 0$, where:
\begin{equation}
    b_0 = \frac{1}{2} (\left\lVert\bm{p}_2\right\rVert_2^2 - \left\lVert\bm{p}_1\right\rVert_2^2)
\end{equation}
This $b_0$ represents the offset of the hyperplane from the origin.

\subsection{Signed Orthogonal Distance to a Hyperplane}
\label{appendix:signed_orthogonal_distance}
For a hyperplane defined by $\bm{w}^\top\bm{x} - b_0 = 0$, the signed orthogonal distance from an arbitrary point $\bm{x}_i$ to this hyperplane is given by:
\begin{equation}
    d_i = \frac{\bm{w}^\top \bm{x}_i - b_0}{\left\lVert\bm{w}\right\rVert_2}
\end{equation}
This formula corresponds to the calculation used in \autoref{eq:signed_distances} and is implemented in \autoref{alg:bipartition2}.

\subsection{Shifting the Hyperplane for Balanced Partitioning}
\label{appendix}
After identifying an optimal anchor pair $(\bm{p}_1, \bm{p}_2)$ and calculating the signed distances $d_i$ for all points in $S$, the median of these distances $\Delta d = \mathrm{Median}(\mathcal{D}_{\text{dist}})$ is used to fine-tune the hyperplane's position. The initial hyperplane is $\bm{w}^\top\bm{x} - b_0 = 0$. We want to shift this hyperplane by $\Delta d$ along its normal vector $\bm{w}$. A point $\bm{x}$ lies on the new, shifted hyperplane if its signed distance to the \textit{original} hyperplane is $\Delta d$. That is:
\begin{equation}
    \frac{\bm{w}^\top \bm{x} - b_0}{\left\lVert\bm{w}\right\rVert_2} = \Delta d
\end{equation}
Rearranging this, we get:
\begin{equation}
    \bm{w}^\top \bm{x} - b_0 = \Delta d \cdot \left\lVert\bm{w}\right\rVert_2
\end{equation}
\begin{equation}
    \bm{w}^\top \bm{x} - (b_0 + \Delta d \cdot \left\lVert\bm{w}\right\rVert_2) = 0
\end{equation}
Thus, the new offset $b$ for the fine-tuned hyperplane is:
\begin{equation}
    b = b_0 + \Delta d \cdot \left\lVert\bm{w}\right\rVert_2
\end{equation}
Substituting the expression for $b_0$:
\begin{equation}
    b = \frac{1}{2} (\left\lVert\bm{p}_2\right\rVert_2^2 - \left\lVert\bm{p}_1\right\rVert_2^2) + \Delta d \cdot \left\lVert\bm{w}\right\rVert_2 \label{eq:final_offset}
\end{equation}
This equation is used in \autoref{alg:build} (line 7) to determine the final hyperplane equation for partitioning. This derivation confirms the dimensional consistency and geometric meaning of the formulas used.

\end{document}